\documentclass[aps,prd,superscriptaddress,nofootinbib,eqsecnum,twocolumn]{revtex4-1}

\pdfoutput=1

\usepackage{graphicx}  % needed for figures
\usepackage{dcolumn}   % needed for some tables
\usepackage{xcolor}
\usepackage{graphics} 
\usepackage{hyperref}
\usepackage{amsfonts}
\usepackage{latexsym}
\usepackage{slashed}
\usepackage{bm}  % To make greek letters in bold face {\bm \Phi}
\usepackage[normalem]{ulem} %tabela
\usepackage{amsmath,mathtools,mathrsfs,bbm,amssymb} 
\usepackage{bbold}
\usepackage{subfigure}
\usepackage{array}
\usepackage{wrapfig}
\usepackage{multirow}
\usepackage{tabularx}
\usepackage{float}

\begin{document}

\title{Charged scalars at finite electric field and temperature in the optimized perturbation theory}

\author{William R. Tavares} \email{tavares.william@ce.uerj.br}
\affiliation{Departamento de F\'{\i}sica Te\'orica, Universidade do Estado do Rio de Janeiro, 
20550-013 Rio de Janeiro, RJ, Brazil} 

\author{Rudnei O. Ramos}
\email{rudnei@uerj.br}
\affiliation{Departamento de F\'{\i}sica Te\'orica, Universidade do Estado do Rio de Janeiro, 
20550-013 Rio de Janeiro, RJ, Brazil} 

\author{Ricardo L. S. Farias}
\email{ricardo.farias@ufsm.br}
\affiliation{Departamento de F\'{\i}sica, Universidade Federal de Santa
  Maria, 97105-900 Santa Maria, RS, Brazil}

\author{Sidney S. Avancini}
\email{sidney.avancini@ufsc.br}
\affiliation{Departamento de F\'{\i}sica, Universidade Federal de Santa
  Catarina, 88040-900 Florian\'{o}polis, SC, Brazil}

%%%%%%%%%%%%%%%%%%%%%%%%%%%%%%%%%%%%%%%%%%%%%%%%% 
\begin{abstract}

We study the symmetry breaking and restoration behavior of a self-interacting 
charged scalar field theory under the influence of a constant electric 
field and finite temperature. Our study is performed in the context
of the optimized perturbation theory. The dependence of the effective
potential with constant electric fields is established by means of the
bosonic propagators in the Schwinger proper-time method.  Explicit
analytical expressions for the electric and thermal contributions are
found. Our results show a very weak decreasing behavior of the vacuum expectation value as a function of the electric field, which is strengthened by the temperature effect. A first-order phase transition that occurs at zero /weak electric fields changes to a second-order phase transition under strong electric fields. The critical temperature for the phase transition exhibited 
a very weak dependence on the electric field. Additionally, we computed the 
vacuum persistence probability rate for the interacting theory, finding a 
peak at the critical point. The maximum value of this rate at the critical 
point is found to be independent of the coupling constant but depended 
solely on the magnitude of the electric field.

\end{abstract}
%%%%%%%%%%%%%%%%%%%%%%%%%%%%%%%%%%%%%%%%%%%%%%%%% 

\maketitle

%%%%%%%%%%%%%%%%%%%%%%%%%%%%%%%%%%%%%%%%%%%%%%%%% 
\section{Introduction}

The study of physical systems through their symmetry aspects 
is a fundamental approach in modern physics. By defining an 
order parameter, we can investigate the different phases exhibited 
by a system~\cite{Goldenfeld:1992qy,Linde:1978px}. Understanding 
how these systems behave under the influence 
of external parameters, such as temperature, density, magnetic fields, 
or electric fields, is crucial for describing realistic scenarios with 
applications in fields like cosmology, condensed matter, and particle physics.

While the effects of magnetic fields on phase transitions have been 
extensively studied (see, for instance,
Refs.~\cite{Miransky:2015ava,Andersen:2014xxa,Hattori:2023egw,Endrodi:2024cqn} for
reviews and references therein), research on the influence of electric fields is 
less common. Electric fields have significant applications in condensed matter systems, 
such as in the case of assisted Schwinger pair production~\cite{Allor:2007ei}. 
This phenomenon occurs when an electric field aids in the creation of 
particle-antiparticle pairs near the Fermi surface. An example of this is
graphene layers, which due to their unique electronic properties, offer opportunities 
to study similar vacuum structures as those found in quantum 
electrodynamics~\cite{Akal:2018txb}.

Much of the existing literature has focused on the effects of electric fields 
on the quantum chromodynamics (QCD) phase transition. Strong electromagnetic 
fields can be produced in peripheral heavy ion 
collisions~\cite{Deng:2012pc,Bzdak:2011yy,Bloczynski:2012en,Bloczynski:2013mca} 
and in asymmetric collisions~\cite{Hirono:2012rt,Voronyuk:2014rna,Deng:2014uja}. 
External magnetic fields play a significant role in heavy ion collisions at typical energies 
(e.g., at those energies achieved in the RHIC and ALICE colliders), where intensities 
of the order $eB\sim10^{19}$G can be achieved~\cite{Huang:2015oca}. Similar 
to magnetic fields, electric fields can serve as an interesting control parameter 
to study important aspects of QCD. Recent numerical analyses have been employing 
intermediate energy per nucleon pair, e.g., $\sqrt{s_{NN}}=3 - 20 \text{ GeV}$, 
in heavy ion collisions. These studies demonstrate that, due to baryon stopping, 
these energies can indeed play a crucial role in creating electric fields that 
persist long enough to modify important hadronic processes~\cite{Panda:2024ccj,Taya:2024wrm}. 
Electromagnetic fields present in heavy ion collisions are also expected to 
influence the dynamics of quark-gluon plasma, such as through anomalous 
transport phenomena like the chiral magnetic effect~\cite{Fukushima:2008xe} 
and chiral separation effect~\cite{Huang:2015oca,Son:2004tq}.

In the context of application to QCD, it is expected that electric
fields can suppress the chiral
condensate~\cite{Klevansky:1992qe,Klevansky:1989vi}, even in the case
of chromo-electric fields~\cite{Suganuma:1991ha}. Also, chiral
perturbation theory has some results concerning to the properties of
nucleons and its decays in the weak field
regime~\cite{Tiburzi:2008ma}, but just only recently is that electric
fields have been combined with temperatures in effective models, as
the Nambu--Jona-Lasinio and its
extensions~\cite{Tavares:2018poq,Tavares:2019mvq,Ruggieri:2016xww,Ruggieri:2016lrn,Correa:2023ebh},
the linear sigma model coupled with quarks~\cite{Tavares:2023ybt} and
Dyson-Schwinger equations~\cite{Ahmad:2020ifp}. Those studies have
indicated that the combined effects of temperature and electric fields
tend to strengthen the restoration of the partial chiral symmetry,
accompanied by the decrease of the pseudocritical temperature until
$eE\sim0.3$ GeV$^2$, where it starts to increase. Studies of the 
deconfinement transition have shown similar qualitative behavior 
to the chiral transition's pseudocritical temperature~\cite{Tavares:2023ybt}. 
Lattice approaches, on the other
hand, have been applied in the case of imaginary electric
fields~\cite{Yang:2023zzx}. However, recent new techniques have been
used in order to avoid technical issues similar to the sign problem,
such as the use of isospin electric charge in the Minkowskian electric
field~\cite{Yamamoto:2012bd}. New research indicates that the deconfinement 
temperature increases with electric field strength~\cite{Endrodi:2023wwf}, 
suggesting the need for refinements in model approaches.
 
While there have been attempts to study strongly interacting quark matter 
using chiral and deconfinement phase transitions in strong electric fields, 
there's a lack of research on charged scalar theories. Recent predictions for
$\lambda\phi^4$ theories have been made using one-loop bosonic propagators in 
finite spatial confining systems~\cite{Correa:2024asi} and through ring diagrams 
at low and high temperatures~\cite{Loewe:2021mdo,Loewe:2022aaw,Loewe:2022plp}.
In the present paper, we use the method of the optimized perturbation theory
(OPT)~\cite{Okopinska:1987hp,Duncan:1988hw} (see also, e.g.,
Ref.~\cite{Yukalov:2019nhu} for a recent review) to study the phase
transition for a self-interacting complex scalar field model at finite
temperature and in a constant electric field.  The OPT is a
nonperturbative method able to resum loop contributions in a
self-consistent way.  It has been so far applied to a large variety of
problems, ranging from condensed matter
systems~\cite{deSouzaCruz:2000fy,Caldas:2008zz,Caldas:2009zz,Gomes:2023vvu},
chiral phase transition in QCD effective
models~\cite{Kneur:2010yv,Kneur:2012qp,Restrepo:2014fna,Duarte:2015ppa}
and in quantum field theory in
general~\cite{Klimenko:1992av,Pinto:1999py,Pinto:1999pg,Kneur:2007vj,Kneur:2007vm,Farias:2008fs,Farias:2021ult,Silva:2023jrk,Martins:2024dag}. 
It has been demonstrated that it has a fast
convergence~\cite{Kneur:2002kq,Rosa:2016czs}, which is of particular
importance, since already at first order the OPT is able to produce
results improving over other nonperturbative methods, e.g., the
large-$N$ expansion and Gaussian approximations, becoming equivalent
to the daisy and superdaisy nonperturbative schemes.

We extend the analysis originally conducted in Ref.~\cite{Duarte:2011ph}, 
which studied the complex scalar field model but under the influence of an external magnetic 
field. Our findings demonstrate that the effects of electric fields on the 
symmetry properties of this model are not only qualitatively different but 
also quantitatively distinct compared to the effects of magnetic fields. 
The methodology we employed to incorporate the electric field allows us 
to investigate the full range of phase transitions in terms of the electric field 
and temperatures. In addition to studying the phase transition, we explore the 
vacuum persistence probability rate (also known as the Schwinger's rate 
for charged particle creation~\cite{Schwinger:1951nm,Schwinger:1954zza,Schwinger:1954zz}), 
which is derived from the imaginary part of the thermodynamic potential. 
We discovered that the vacuum persistence probability rate is enhanced at 
the critical point of the phase transition. This could potentially serve as 
a valuable indicator of the critical point in physical systems of interest. 
To our knowledge, this is the first study of its kind to be conducted in 
the context of an interacting scalar field theory and across the phase transition.

This work is organized as follows. In Sec.~\ref{sec2}, we introduce
the model and detail its OPT implementation to obtain the
effective potential. In Sec.~\ref{sec3}, the temperature and electric
field are introduced through the Schwinger's proper time approach and
the scalar field propagator for the model is given. The effective
potential at first order in the OPT is explicitly computed. In this
same section, we also define and give the result for the vacuum
persistent probability rate at the same order of approximation in the
OPT scheme.  In Sec.~\ref{sec4}, we present our main results. The
effective potential is studied for different values of temperature and
electric field and the phase behavior of the system is explicitly
analyzed. The results for the vacuum persistence probability rate are
also presented and studied how it is affected by temperature and for
different magnitudes for the electric field. Our conclusions are
presented in Sec.~\ref{sec5}. {}Four appendices are included where the
technical aspects and details of the derivations are presented.

Throughout this paper, we work with the natural units, in which the
speed of light,  Planck's constant and Boltzmann's constant are all
set to $1$, $c=\hbar=k_B=1$. We take the Minkowski metric to be ${\rm diag}(+,-,-,-)$.

%%%%%%%%%%%%%%%%%%%%%%%%%%%%%%%%%%%%%%%%%%%%%%%%% 
\section{The model and the OPT implementation}
\label{sec2}

We start by considering the simple case of the self-interacting
$\lambda \phi^4$ model, with Lagrangian density given by
\begin{eqnarray}
\mathcal{L}=(D_{\mu}\phi)^{\dagger}(D^{\mu}\phi)-V(\phi \phi^\dagger),
\label{lagrangian}
\end{eqnarray}
with  a spontaneous symmetry breaking potential ($m^2 >0$),
\begin{equation}
V(\phi \phi^\dagger) = - m^2 \phi \phi^\dagger + \frac{\lambda}{3!}
(\phi \phi^\dagger)^2,
\label{Vphi}
\end{equation}
and where  $D_\mu = \partial_\mu + i q A_\mu$, with $A_\mu$ the
Abelian gauge field, here considered as an external background field.  As usual,
writing the complex field $\phi$ as
$\phi=(\phi_1+i\phi_2)/\sqrt{2}$. As usual, we can choose to shift the
field around the $\phi_1$ direction, $\phi_1\to \phi_1+\varphi$, where
$\varphi$ is a background field and which the effective potential is a
function of. Then, the symmetry breaking potential is now given by

\begin{eqnarray}
    V(\varphi,\phi_1,\phi_2)&=&-\frac{m^2}{2}[(\phi_1+\varphi)^2+\phi_2^2]\nonumber\\&&+\frac{\lambda}{4!}[(\phi_1+\varphi)^2+\phi_2^2]^2.
\end{eqnarray}

When $\varphi$ minimizes the potential, $\varphi \equiv
\bar \varphi$, by applying

\begin{eqnarray}
    \frac{\partial V(\phi_1,\phi_2,\varphi)}{\partial\phi_1}\Bigr|_{\phi_1=\bar{\varphi},\phi_2=0,\varphi=\bar{\varphi}}=0,
\end{eqnarray}

\noindent it becomes the vacuum expectation  value (VEV) of the
scalar field, $|\langle \phi_1 \rangle|=\bar\varphi$ and $|\langle
\phi_2 \rangle|=0$.  {}For the tree-level potential, we have from
Eq.~(\ref{Vphi}) that $\bar\varphi\equiv\varphi_0=\pm
\sqrt{6m^2/\lambda}$.  Hence, in the symmetry broken phase $\phi_1$ is
associated with the Higgs field, while $\phi_2$ is the Goldstone
field.  Their {}Feynman propagators, written in terms of the VEV, is
given by
\begin{eqnarray}
D_{\phi_1}(p)=\frac{i}{p^2+m^2-\frac{\lambda}{2}\varphi^2+i\epsilon},
\label{D1}\\
D_{\phi_2}(p)=\frac{i}{p^2+m^2-\frac{\lambda}{6}\varphi^2+i\epsilon}.
\label{D2}
\end{eqnarray}

The OPT implementation starts by interpolating the model
(\ref{lagrangian}) such that~\cite{Duarte:2011ph}
\begin{eqnarray}
  \mathcal{L}\rightarrow\mathcal{L_{\delta}}&=&\sum_{i=1}^2
  \left\{\frac{1}{2}(\partial_{\mu}\phi_i)^2-
  \frac{1}{2}\Omega^2\phi^2_i+\frac{\delta}{2}\eta^2\phi^2_i-
  \delta\frac{\lambda}{4!}\phi_i^4
\right\}\nonumber\\ &+&\mathcal{L}_{ct,\delta},
\label{LagOPT}
\end{eqnarray}
with $\Omega^2= -m^2 + \eta^2$ and where $\mathcal{L}_{ct,\delta}$ is
the renormalization counterterm contribution to the Lagrangian
density.  In the interpolated Lagrangian density Eq.~(\ref{LagOPT}),
$\eta$ is a mass term introduced by the OPT procedure and $\delta$ is
a bookkeeping (dimensionless) parameter controlling the order in which
the OPT method is carried out. The mass parameter $\eta$ is obtained
at each order of the OPT approximation through a variational
principle, which in here we use the principle of minimal sensitivity
(PMS)~\cite{Stevenson:1981vj}\footnote{Note that are also other
possible alternative optimization criteria that can be used, similar
to the PMS one~\cite{Farias:2008fs,Rosa:2016czs} and that produces
equivalent results. In the present paper we will not consider those
alternative procedures.}. The PMS criterion consists of applying the
variational condition to some physical quantity under consideration,
which here is taken to be the effective potential, $V_{\rm eff}$. The
PMS criterion then produces an optimal variational mass
$\overline{\eta}$, which can then be obtained by the relation
\begin{equation}
\left.\frac{d V_{\rm eff}}{d
  \eta}\right|_{\bar{\varphi},\bar{\eta},\delta=1}=0,
\label{PMS}
\end{equation}
while the VEV for the scalar field, $\bar \varphi$, is obtained by
minimizing the effective potential,
\begin{eqnarray}
    \frac{dV_{\rm
        eff}}{d\varphi}\bigg\rvert_{\bar{\varphi},\bar{\eta},\delta=1}=0.
\label{vev}
\end{eqnarray}

We can derive the Feynman rules by the interpolated Lagrangian density in Eq.~(\ref{LagOPT}) by changing the vertex $-i\lambda$ by $-i\delta\lambda$; the new term $\delta\eta^2\phi_i^2/2$ is interpreted as a new vertex in the OPT formalism. In terms of the interpolated model, the propagators given by
Eqs.~(\ref{D1}) and (\ref{D2}) now become
\begin{eqnarray}
  D_{\phi_1,\delta}(p)=\frac{i}{p^2-\Omega^2-\frac{\lambda\delta}{2}\varphi^2+
    i\epsilon},
\label{prop1}
\\
  D_{\phi_2,\delta}(p)=\frac{i}{p^2-\Omega^2-\frac{\lambda\delta}{6}\varphi^2+
    i\epsilon}.
\label{prop2}
\end{eqnarray}

\begin{center}
\begin{figure}[!htb]
\includegraphics[width=7.5cm]{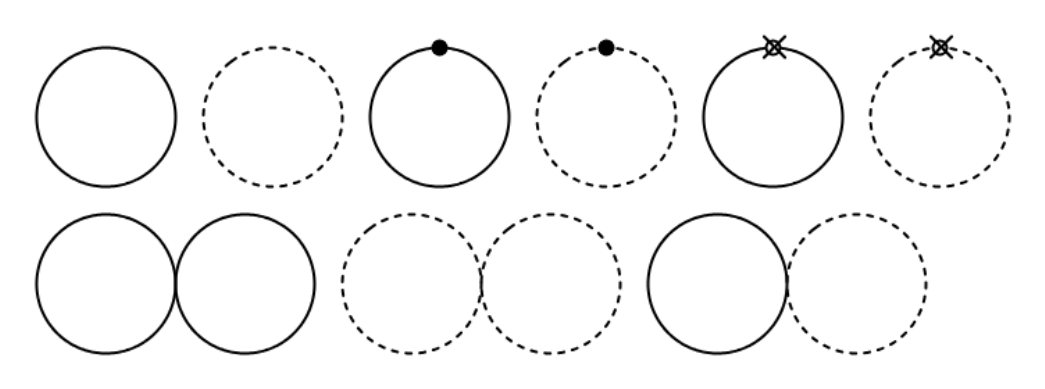}\\
\caption{Vacuum Feynman diagrams that contribute to order $\delta$ to the effective potential. Solid lines corresponds to the propagator of $\phi_1$, and dashed lines corresponds to $\phi_2$. The black dots corresponds to $\delta\eta^2$ insertions, while the vacuum diagrams with crosses are represents the counter terms. }
\label{fig1}
\end{figure}
\end{center}

The vacuum diagrams shown in {}Fig.~\ref{fig1}, represent the effective potential contributions to order $\delta$ in the OPT.

%%%%%%%%%%%%%%%%%%%%%%%%%%%%%%%%%%%%%%%%%%%%%%%%%%%
%%%%%%%%%%%%%%%%%%%%%%%%%%%%%%%%%%%%%%%%%%%%%%%%%%%

 By inserting the propagators (\ref{prop1}) and (\ref{prop2}) 
in the vacuum diagrams shown in {}Fig.~\ref{fig1} and expanding again in $\delta$, the effective potential at first order in the OPT
is obtained to be given by~\cite{Duarte:2011ph}
\begin{eqnarray}
V_{\rm
  eff}&=&\frac{\Omega^2}{2}\varphi^2-\delta\frac{\eta^2}{2}\varphi^2+
\delta\frac{\lambda}{4!}\varphi^4
\nonumber \\ &-&\!\!\!  i\int_p\ln\left(p^2-\Omega^2+i\epsilon
\right)-\delta\eta^2\int_p\frac{i}{p^2-\Omega^2+i\epsilon} \nonumber
\\ &+&\!\!\!\delta\frac{\lambda}{3}\varphi^2\!\!\int_p\frac{i}{p^2-\Omega^2+
  i\epsilon}
+\delta\frac{\lambda}{3}\left(\int_p\frac{i}{p^2-\Omega^2+i\epsilon}\right)^2,
\nonumber \\
\label{Veff}
\end{eqnarray}
where we have defined $\int_p\equiv\int \frac{d^4p}{(2\pi)^4}$. It is
convenient to define an effective  scalar propagator in the OPT as
\begin{eqnarray}
    D_{\Omega}(p)=\frac{i}{p^2-\Omega^2+i\epsilon},
\label{DOPT}
\end{eqnarray}
and also to express the momentum integrals in the effective potential
(\ref{Veff}), using  dimensional regularization in the $\overline{\rm
  MS}$-scheme, as
\begin{eqnarray}
&& \int_p D_{\Omega}(p) =  -\frac{\Omega^2}{(4\pi)^2}
  \frac{1}{\epsilon} + X(\eta),
\label{DX}
\\ && \!\!\!\!\!\!\!\!\!
-i\int_P\ln\left(-iD_{\Omega}(p)^{-1}\right)=- \frac{\Omega^4}{2 (4
  \pi)^2} \frac{1}{\epsilon} + Y(\eta),
\label{DY}
\end{eqnarray}
where $X(\eta)$ and $Y(\eta)$ are functions of
the temperature $T$ and electric field $E$, and they will be explicitly defined
in the next section. In the vacuum, $T=0$ and $E=0$, they are given by
\begin{eqnarray}
  &&X(\eta)\Bigr|_{T=0,E=0}=\frac{\Omega^2}{16\pi^2}
  \left[\ln\left(\frac{\Omega^2}{M^2}\right)-1\right],
\label{X0}
\\ &&Y(\eta)\Bigr|_{T=0,E=0}=-\frac{\Omega^4}{2(4\pi)^2}
\left[\frac{3}{2}-\ln\left(\frac{\Omega^2}{M^2}\right)\right].
\label{Y0}
\end{eqnarray}
Thus, by adding the appropriate counterterms of renormalization to
cancel the ultraviolet divergences in (\ref{Veff}), the renormalized
effective potential in the first-order approximation in the OPT can be
generically written as
\begin{eqnarray}
    V_{\rm
      eff}(\varphi,\eta)&=&-\frac{m^2}{2}\varphi^2+(1-\delta)
    \frac{\eta^2}{2}\varphi^2+
    \delta\frac{\lambda}{4!}\varphi^4+Y(\eta)
    \nonumber\\ &+&\delta\left\{-\eta^2+\frac{\lambda}{3}
    \left[\varphi^2+X(\eta)\right]\right\}
    X(\eta),
\label{VeffR}
\end{eqnarray}
and using the PMS condition (\ref{PMS}) and from the VEV equation
(\ref{vev}), we have that
\begin{eqnarray}
  \bar{\eta}^2=\frac{\lambda}{3}\bar\varphi^2+
  \frac{2\lambda}{3}X(\bar{\eta}),
\label{bareta}
\\ \bar{\varphi}^2=6\frac{m^2}{\lambda}-4X(\bar{\eta}).
\label{barvev}
\end{eqnarray}
The effective potential Eq.~(\ref{VeffR}), with the
Eqs.~(\ref{bareta}) and (\ref{barvev}) form the basic equations that
we need to completely explore the phase structure of the model.

%%%%%%%%%%%%%%%%%%%%%%%%%%%%%%%%%%%%%%%%%%%%%%%%% 
\section{The effective potential at finite temperature and in a constant
  electric field}
\label{sec3}

Let us now explicitly introduce the effects of temperature and electric
field.  We start by defining the electric field dependent bosonic
propagator, $D_{\Omega}(P,qF)$,
by~\cite{Ahmad:2016vvw,Edwards:2017bte,Hattori:2023egw}
\begin{eqnarray}
D_{\Omega}(p,qF)=i\int_{0}^{\infty}
ds\frac{e^{-is\Omega^2}e^{ip_\mu\left[\frac{\tanh
        (qFs)}{qF}\right]^{\mu \nu}p_\nu}}{\sqrt{ \det \left[
      \cosh(qFs)\right] }},
\end{eqnarray}
where $F_{\mu\nu}$ is the electromagnetic field tensor. The
electric and magnetic fields are expressed by $E^i=F^{i0}$ and $B^i =
-\sum_{j,k} \varepsilon^{ijk} F^{jk}/2$. Hence, we can set for definiteness the only nonzero fields to $i=3$ in both cases in order to have the bosonic propagator in a constant parallel magnetic and electric fields oriented in the $z$ direction, to obtain
\begin{eqnarray}
D_{\Omega}(p,qF)=i\int_{0}^{\infty}
ds\frac{e^{-is\Omega^2}e^{i\frac{\tanh (qEs)}{qE}
    p_{\parallel}^2+i\frac{\tan(qBs)}{qB} p_{\perp}^2
}}{\cosh(qEs)\cos(qBs) },\nonumber \\
\label{DEB}
\end{eqnarray}
where $p_{\perp}^2=-(p_1^2+p_2^2)$ and
$p_{\parallel}^2=p_0^2-p_3^2$. In the present paper, we work in the
case of a non-null constant electric field, $E\neq 0$, and in the absence of a
magnetic field, $B=0$. It is also more convenient to define the
expressions in the Euclidean spacetime. Thus, with $s\to -i\tau$,
$p_0\to i p_4$ and $B=0$, the (Euclidean) scalar propagator becomes
\begin{eqnarray}
D_{{\rm Eucl},\Omega}(p,qE)=\int_{0}^{\infty} d\tau\frac{e^{-\tau
    \Omega^2}\;e^{\left(\frac{\tan
      (qE\tau)}{qE\tau}p_{\parallel}^2+p_{\perp}^2\right)\tau}}{\cos(qE\tau)},
\end{eqnarray}
where, in Euclidean spacetime, $p_{\parallel}^2=-(p_4^2+p_3^2)$.

{}Finite temperature is included through the Matsubara imaginary-time
formalism, where $p_4=\omega_n$, $\omega_n=2n\pi T$ are the bosonic
Matsubara frequencies with $n=0,\pm1,\pm2,\ldots$. Here, we also work
in the dimensional regularization method to perform the
momentum integrals, which are given by
\begin{eqnarray}
\int_p \equiv iT \sum_{p_0=i\omega_n} \left(\frac{e^{\gamma_E}
  M^2}{4\pi}\right)^\epsilon \int\frac{d^{d}p}{(2\pi)^d}\;,
\label{intT}
\end{eqnarray}
where $M$ is the regularization scale (in the $\overline{\rm
  MS}$-scheme), $\gamma_E$ is the Euler-Mascheroni constant and $d=
3-\epsilon$  in dimensional regularization. This approach will give
raise to a thermoelectric contribution, obtained in an analogous way
as in the fermionic case~\cite{Tavares:2023ybt}.  Proceeding now with
the evaluation of the renormalized expression for $X(\eta,T,qE)$ and
adopting the change of variables $qE\tau\rightarrow t$, we have that
\begin{eqnarray}
    X(\eta,T,qE)&=&
    \frac{\Omega^2}{16\pi^2}\left[\ln\left(\frac{\Omega^2}{M^2}\right)-1\right]
    \nonumber \\ &+&
    \frac{qE}{16\pi^2}\int_0^\infty\frac{dt}{t}e^{-t\frac{\Omega^2}{qE}}
    \left(\frac{1}{\sin(t)}-\frac{1}{t}\right)
    \nonumber\\ &+&\frac{qE}{8\pi^2}\int_0^\infty\frac{dt}{t\sin(t)}
    e^{-t\frac{\Omega^2}{qE}}\sum_{n=1}^{\infty}e^{-\frac{n^2qE}{4T^2|\tan(t)|}}.
    \nonumber \\
\label{XE}
\end{eqnarray}
Likewise, for $Y(\eta,T,qE)$ we have that
\begin{eqnarray}
Y(\eta,T,qE)&=&
-\frac{\Omega^4}{32\pi^2}\left[\frac{3}{2}-\ln\left(\frac{\Omega^2}{M^2}
  \right)\right]
\nonumber
\\   &+&\frac{(qE)^2}{16\pi^2}\int_0^\infty\frac{dt}{t^2}
e^{-t\frac{\Omega^2}{qE}}\left(\frac{1}{\sin(t)}-\frac{1}{t}-\frac{t}{6}\right)
\nonumber\\ &-&\frac{(qE)^2}{8\pi^2}\int_0^\infty\frac{dt}{t^2\sin(t)}
e^{-t\frac{\Omega^2}{qE}}\sum_{n=1}^{\infty}e^{-\frac{n^2qE}{4T^2
    |\tan(t)|}}.  \nonumber\\
\label{YE}
\end{eqnarray}

It is clear from  the functions in the Eqs.~(\ref{XE}) and (\ref{YE})
that there are periodic poles, due to the $\sin(t)$ function, when
$t=0,\pi,2\pi,3\pi, \ldots$. To deal with these poles, it is
convenient to separate the real and imaginary parts of the functions
$X(\eta,T,qE)$ and $Y(\eta,T,qE)$ and interpreting the physical
results by means of the real contributions.  Details of this procedure
are given in the Appendices~\ref{appA}, \ref{appB} and \ref{appC}. The
final result for  the real part of the function $X(\eta,T,qE)$ is
given by
\begin{eqnarray}
&&\Re
  X(\eta,T,qE)=\frac{\Omega^2}{16\pi^2}\ln\left(\frac{2qE}{M^2}\right)
  \nonumber\\ && +\frac{qE}{8\pi^2}\Bigl\{-\gamma_E \;y-\arctan(2y)
  \nonumber\\ && + \sum_{n=1}^{\infty}\left[
    \frac{y}{n}-\arctan\left(\frac{2y}{2n+1}\right)\right]\Bigr\}
  \nonumber\\ &&+\frac{qE}{8\pi^2}\int_0^\infty\frac{dt}{t\sin(t)}
  e^{-t\frac{\Omega^2}{qE}}\sum_{n=1}^{\infty}e^{-\frac{n^2qE}{4T^2|\tan(t)|}},
\label{Xfull}
\end{eqnarray}
while for $\Re Y(\eta,T,qE) $, we obtain that
\begin{eqnarray}
&&\Re Y(\eta,T,qE)
  =\frac{\Omega^4}{32\pi^2}\ln\left(\frac{2qE}{M^2}\right)
  \nonumber\\ &&-\frac{(qE)^2}{4\pi^2}\left\{\frac{\gamma_Ey^2}{2}+
  y\arctan(2y)-\frac{1}{4}\ln\left(1+4y^2\right)\right.
  \nonumber\\ &&\left.-\sum_{k=1}^{\infty}\left[\frac{y^2}{2k}-
    y\arctan\left(\frac{2y}{2k+1}\right)
    \right. \right.  \nonumber
    \\ &&+\left.\left. \frac{2k+1}{4}
    \ln\left(1+\frac{y^2}{(2k+1)^2}\right)\right]\right\}
  \nonumber\\
  &&-\frac{(qE)^2}{8\pi^2}\int_0^\infty\frac{dt}{t^2\sin(t)}
  e^{-t\frac{\Omega^2}{qE^2}}\sum_{n=1}^{\infty}e^{-\frac{n^2qE}{4T^2|\tan(t)|}},
  \label{Yfull}
\end{eqnarray}
where, in the above equations, we have defined $y=\Omega^2/(2qE)$.
The thermal integrations in the Eqs.~(\ref{Xfull}) and (\ref{Yfull})
are all convergent and they can be integrated numerically in a similar
way as it was done in Ref.~\cite{Tavares:2018poq}.

The physical renormalized effective potential at first order in the
OPT is then given by the  real part of $V_{\rm eff}$, 
\begin{eqnarray}
&&\Re V_{\rm
    eff}(\varphi,\eta,T,qE)=-\frac{m^2}{2}\varphi^2+
  \frac{\lambda}{4!}\varphi^4
  \nonumber\\ &&- \left(\eta^2-\frac{\lambda}{3}\varphi^2\right)\Re
  X(\eta,T,qE)  \nonumber \\ &&+\frac{\lambda}{3}\left[\Re
    X(\eta,T,qE)\right]^2   - \frac{\lambda}{3}\left[\Im
    X(\eta,T,qE)\right]^2 \nonumber\\ &&+\Re Y(\eta,T,qE),
\label{ReVeff}
\end{eqnarray}
where the imaginary part of the function $X(\eta,T,qE)$, $\Im
X(\eta,T,qE)$, was derived in the App.~\ref{appC} and given by
Eq.~(\ref{ImX}).  {}From Eq.~(\ref{ReVeff}), the conditions for the
PMS, Eq.~(\ref{PMS}), and for the VEV, Eq.~(\ref{vev}) are then given
by
\begin{eqnarray}
  &&\left[-\bar{\eta}^2+\frac{\lambda}{3}\bar{\varphi}^2+
    \frac{2\lambda}{3}\Re
    X(\bar{\eta},T,qE)\right]\frac{\partial} {\partial \eta}\Re
  X(\eta,T,qE)
  \bigg\rvert_{\eta=\bar{\eta}}\nonumber\\ &&-\frac{2\lambda}{6}
  \frac{\partial}{\partial
    \eta} [\Im
    X(\eta,T,qE)]^2\bigg\rvert_{\eta=\bar{\eta}}=0,\label{PMSelectric}\\
  &&\bar{\varphi}^2=6\frac{m^2}{\lambda}-4\Re
  X(\bar{\eta},T,qE).\label{GAPelectric}
\end{eqnarray}

In addition to real part of the effective potential, we also quote the
expression for the imaginary part of the effective potential,
\begin{eqnarray}
\lefteqn{\Im V_{\rm eff}(\varphi,\eta,T,qE)=\Im Y(\eta,qE) }
\nonumber\\ &&-\left[\eta^2-\frac{\lambda}{3}\varphi^2-
  \frac{2\lambda}{3}\Re
  X(\eta,T,qE)\right]\Im X(\eta,qE), \nonumber\\
\label{IMVEFF}
\end{eqnarray}
where the imaginary part of the function $Y$ was derived in the
App.~\ref{appC} and given by Eq.~(\ref{ImY})\footnote{Issues concerning
the derivation of the imaginary part of the loop diagrams in a thermal
bath are discussed in the App.~\ref{appC}.}.  The imaginary part of
the effective potential here, which is entirely due to the presence of
the background electric field, indicates an instability of the vacuum
towards the productions of charged particles due to the electric
field.  It is known now as the 
{\it vacuum persistence probability rate}~\cite{Dunne:2004nc,Cohen:2008wz}, 
$w$, which from
Eq.~(\ref{IMVEFF}),  we can express it as\footnote{ The vacuum
persistence probability itself is $P_{\rm vac}(t) = e^{-w {\cal V}
  t}$, where ${\cal V}$ is the spatial volume of the system and $w$ is
the rate of vacuum decay per unit volume.}
\begin{eqnarray}
    w(qE,T)=-2\Im V_{eff}(\bar{\varphi},\bar{\eta},T,qE) ,
\end{eqnarray}
which for the present problem it is given explicitly as
\begin{eqnarray}
\lefteqn{w(qE,T) = -\frac{(qE)^2}{8\pi^3}\text{Li}_2(-e^{-2\pi
    y})\Bigr|_{\eta=\bar\eta}} \nonumber \\ &&+
\frac{qE}{16\pi^2}\ln\left(1+e^{-2\pi y}\right)
\nonumber \\
&\times&
\left[\eta^2-\frac{\lambda}{3}\bar\varphi^2-\frac{2\lambda}{3}\Re
  X(\eta,T,qE)\right]\Bigr|_{\eta=\bar\eta}, \nonumber\\
\label{wE}
\end{eqnarray}
where $\Re X$ is given by Eq.~(\ref{Xfull}). It can be checked that in
the free (unbroken) theory case, with $\lambda \to 0$, $\Omega^2\to
m^2$, $y\to m^2/(2qE)$, the result given by Eq.~(\ref{wE}) reproduces
the expression for the rate $w$ for the free scalar quantum
electrodynamics (QED) case~\cite{Dunne:2004nc},
\begin{eqnarray}
w(qE,T)\to w_{\rm free}(qE)
&=&-\frac{(qE)^2}{8\pi^3}\text{Li}_2\left(-e^{-\frac{\pi
    m^2}{qE}}\right) \nonumber\\ &=&
\frac{(qE)^2}{8\pi^3}\sum_{k=1}^\infty \frac{(-1)^{k-1} e^{-\frac{\pi
      m^2 k}{qE}}}{k^2}.  \nonumber \\
\label{wfree}
\end{eqnarray}
Note that the result for the free scalar QED is independent of the
temperature. In this way, we can see that the OPT at first order
includes thermal and interaction dependent corrections to the
imaginary part of the effective potential (and hence to $w$), with the
optimum mass parameter $\bar\eta$ determined by the PMS condition
Eq.~(\ref{PMSelectric}) and with $\bar \varphi$ given by
Eq.~(\ref{GAPelectric}). 

Before we close this Section, it is worth to remark that the imaginary
part of the effective potential given by Eq.~(\ref{IMVEFF}) should not
be confused with the situation when the finite temperature effective
potential can become imaginary, in a symmetry broken theory and in
perturbation theory, for some values of the background field. In that
case, the imaginary part of the effective potential can be interpreted
as a decay width of the localized quantum state, used to
perturbatively evaluate the effective potential, to the true vacuum
state~\cite{Weinberg:1987vp}. As far as the derivation of the
effective potential is concerned, we do not have such issues of
imaginary terms as plagued the perturbative calculation. In
particular, the OPT effective mass squared  appearing in the
propagator Eq.~(\ref{DOPT}), $\Omega^2$, remains always positive
definite, with $\bar \eta^2 \geq m^2$. This is a result known from previous
studies in the context of the OPT (see, in particular the
Refs.~\cite{Duarte:2011ph,Silva:2023jrk}). The same remains true in
the present study, as we explicitly verify in the next Section. In
particular, it is important to also remark that the OPT preserves the
Goldstone theorem, as demonstrated in
Refs.~\cite{Duarte:2011ph,Silva:2023jrk}.

%%%%%%%%%%%%%%%%%%%%%%%%%%%%%%%%%%%%%%%%%%%%%%%%% 
\section{Numerical Results}
\label{sec4}

\begin{center}
\begin{figure}[!htb]
\subfigure[]{\includegraphics[width=7.5cm]{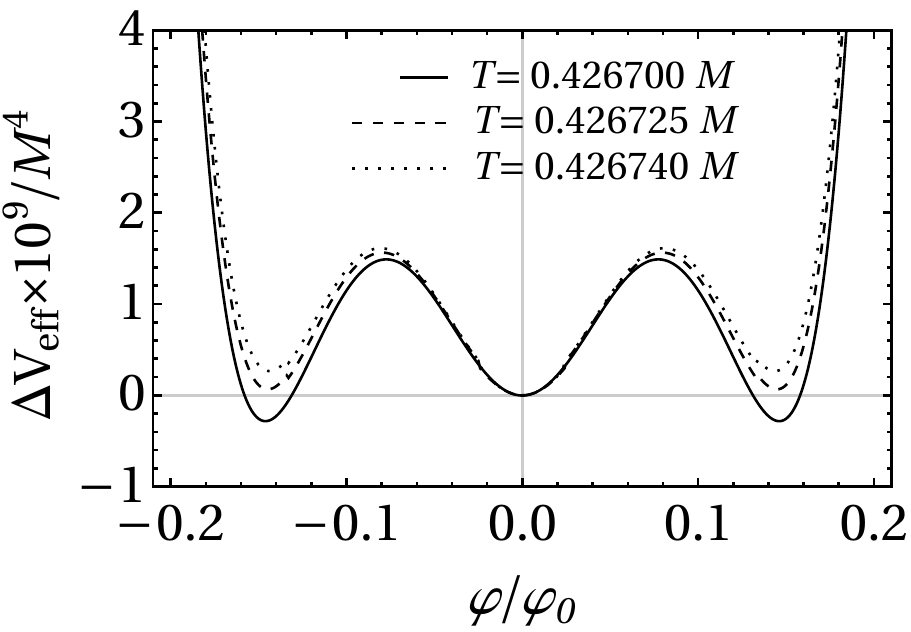}}
\subfigure[]{\includegraphics[width=7.5cm]{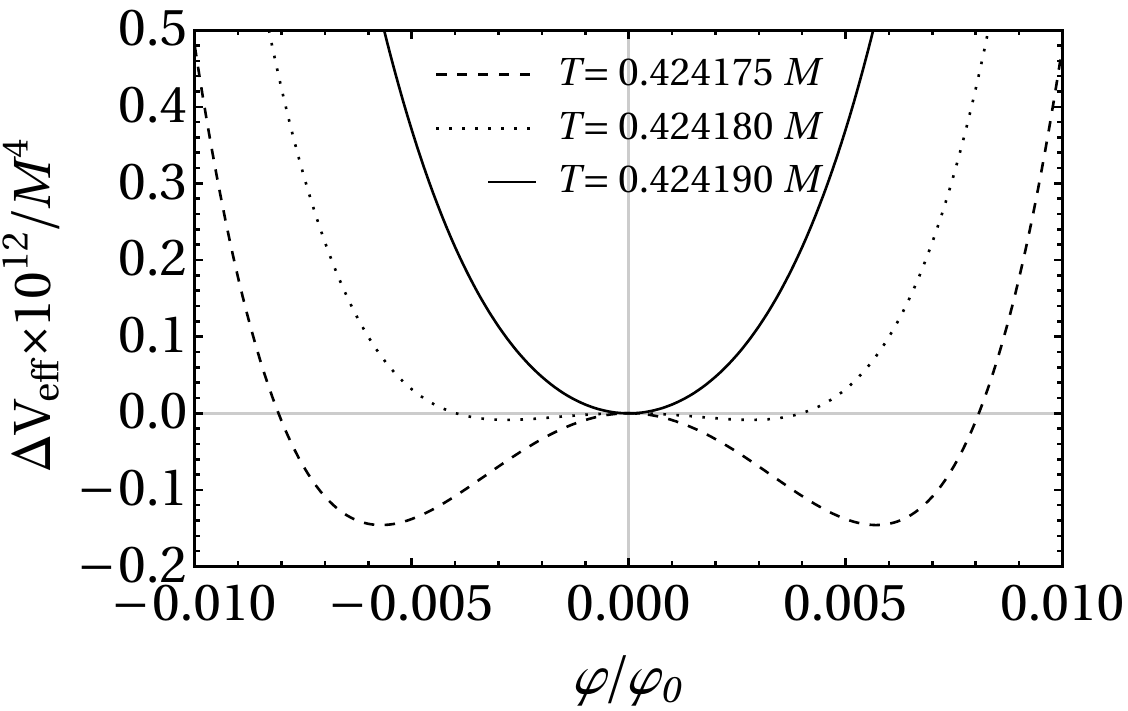}}
\caption{Reduced effective potential $\Delta
  V_{\text{eff}}(qE,T,\bar\eta,\varphi)$ as a function of
  $\varphi/\varphi_0$ for different values of temperature $T/M$ for
  $qE=0$ (panel a) and for $qE=0.2M^2$ (panel b). The model
  parameters chosen are $\lambda=1$ and $m=0.1 M$.}
\label{fig2}
\end{figure}
\end{center}

In this Section, we explore the numerical results of the model.  Let
us start by studying the effective potential Eq.~(\ref{ReVeff}) as a
function of the background field and for different values of
temperature and electric field.  Our results show that there is a
threshold value for the electric field below which the phase
transition is first order, while above it the transition becomes
second order\footnote{ Note that from universality, on expects that the phase transition for a complex scalar field to be of second-order. The result obtained here can in principle be attributed to the order that the OPT approximation is carried out. This is however a point that needs further study. However, note that the transition is very weak first order, where the variation of $T_c$ between $E=0$ and when the transition becomes second-order, obtained by increasing $E$, is very small. Note also that a first-order phase transition for the complex scalar field was also obtained previously in Ref.~\cite{Ayala:2004dx} by using a different approximation method.}.  In {}Fig.~\ref{fig2} we have used a
representative choice of model parameters, e.g., $\lambda=1$ and $m=0.1 M$
(we express all dimensional quantities in terms of the regularization
scale $M$). This is a choice of parameters where the results start
becoming  more sensitive to the effects of the electric field and one
that allows us to compare the results in the absence of $E$ more
easily.

%%%%%%%%%%%%%%%%%%%%%%%%%%%%%%%%%%%%%%%%%%%%%%%%%%%
%%%%%%%%%%%%%%%%%%%%%%%%%%%%%%%%%%%%%%%%%%%%%%%%%%%

We express our results in terms of the reduced effective potential,
$\Delta V_{\text{eff}}=V_{\text{eff}}(\varphi,\bar\eta,T,qE)-
V_{\text{eff}}(0,\bar\eta_0,T,qE)$,
where $\bar\eta_0$ is the optimal value of $\eta$ when evaluated at
$\varphi=0$.  In {}Fig.~\ref{fig2}, we show the reduced effective
potential for different values of temperature around the critical
value.  The transition pattern shown in {}Fig.~\ref{fig2}(a) is that
of a typical first-order phase transition. The electric field
considered in that case was $E=0$. By increasing the electric
field, eventually above a threshold value the transition becomes
second-order. 
 This is illustrated in {}Fig.~\ref{fig2}(b), which
shows a typical second-order phase transition where the VEV decreases
as the temperature increases and becomes zero at the critical
temperature. 
The electric field considered in this case was
$qE=0.2M^2$. The critical temperature for the first-order phase
transition seen in the case shown in {}Fig.~\ref{fig2}(a) is found to
be given by $T_c \simeq 0.426725 M$, while for the case of the
second-order phase transition seen in {}Fig.~\ref{fig2}(b) is $T_c
\simeq 0.424180 M$.

Despite the different transition patterns and changes in the electric 
field, the critical temperature for phase transition remains relatively constant, 
indicating a very weak dependence on the external electric field. 
This intriguing result is further explored below. Previous studies of the 
same model in the absence of electric fields ($E=0$)~\cite{Ayala:2004dx} 
also found a first-order phase transition. In Ref.~\cite{Ayala:2004dx}, 
the authors investigated the phase transition behavior of the complex scalar 
field using the ring diagram approximation. Our results align with this 
finding, as we also observe a first-order phase transition up to 
the threshold value of the electric field within the context of the 
OPT. While not explicitly shown here, we have explored the phase 
transition for different parameter values and found that the threshold 
value for the electric field depends on these choices. 
{}For example, with $\lambda=2$ and $m=M$, we obtain $qE_{\rm threshold} \simeq 0.08 M^2$,  while for $\lambda=1$ and $m=M$ we find that $qE_{\rm threshold} \gtrsim 0.048 M^2$.
In general, increasing the self-coupling tends to increase the threshold value 
for the electric field at which the phase transition changes from first-order 
to second-order.

%%%%%%%%%%%%%%%%%%%%%%%%%%%%%%%%%%%%%%%%%%%%%%%%%%%
\begin{center}
\begin{figure}[!htb]
\subfigure[]{\includegraphics[width=7.5cm]{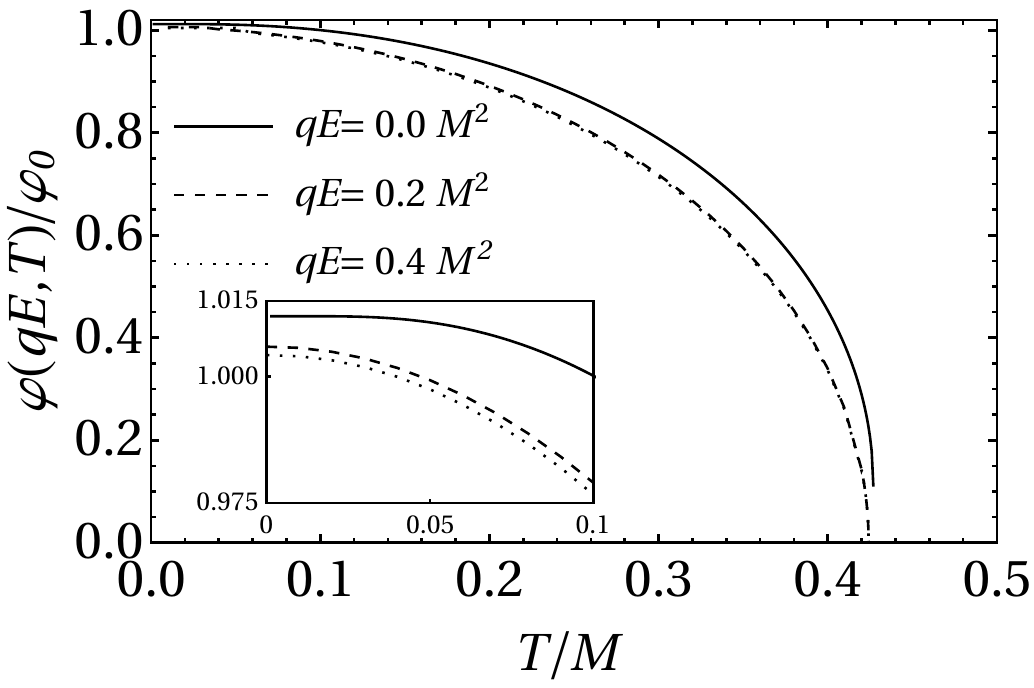}}
\subfigure[]{\includegraphics[width=7.5cm]{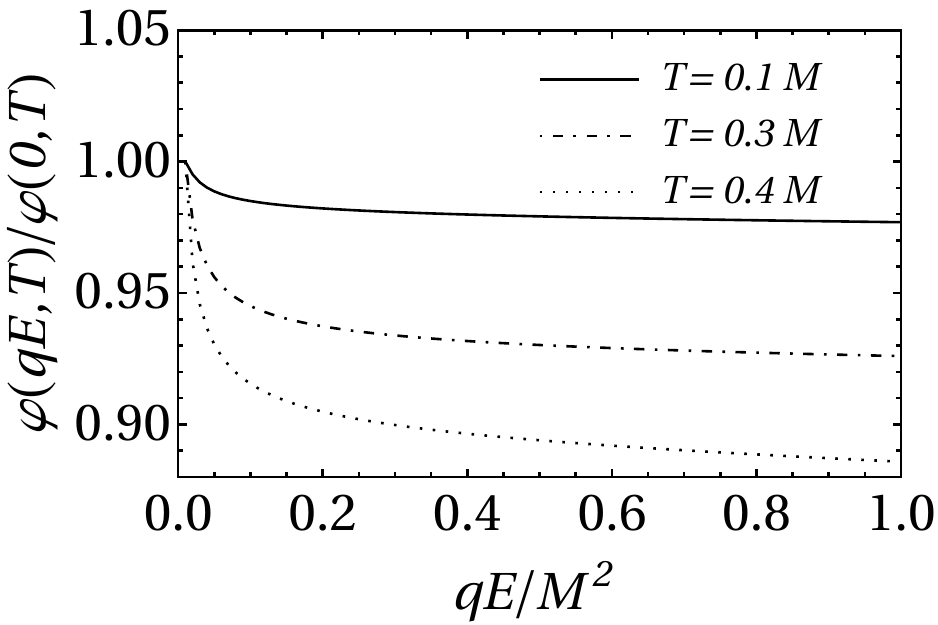}}
\caption{The VEV as a function of the temperature (panel a) and as a
  function of the electric field (panel b). The model parameters chosen are
  $\lambda=1$ and $m=0.1 M$. The inset in panel (a) helps to see the behavior in the region close to $T=0$.}
\label{fig3}
\end{figure}
\end{center}
%%%%%%%%%%%%%%%%%%%%%%%%%%%%%%%%%%%%%%%%%%%%%%%%%%%

It is also relevant to look at the behavior of the VEV as a function
of temperature and electric field. In {}Fig.~\ref{fig3} we show the
normalized VEV $\bar\varphi/\varphi_0$,  where
$\bar\varphi\equiv\varphi(qE,T)$ and $\varphi_0^2=6m^2/\lambda$, which
is shown as a function of the temperature and for different values of
the external electric field (panel a) and as a function of the
electric field and different values of the temperature (panel b). In
{}Fig.~\ref{fig3}(a), we can see that at $qE=0$ we have the usual
symmetry restoration pattern, e.g. as previously observed
in~\cite{Duarte:2011ph}, in which the $\bar\varphi/\varphi_0$
decreases as a function of $T$. Note also the discontinuity of the VEV
close to the critical temperature, indicating a first-order phase
transition, which persists up to the threshold value for the electric
field discussed above. At finite values of electric field considered, we can see that the different curves of $\bar\varphi/\varphi_0$ almost coincide at $T=0$, indicating a very-weak electric field dependence of the VEV at low temperatures. Above the threshold value for $E$, the VEV
always decreases in a smooth way, vanishing at the critical point, as
expected for a second-order phase transition  Note also that for
higher values of $qE$ the magnitude of $\varphi/\varphi_0$ at
intermediate values of temperature in between $T=0$ and $T=T_c$ cause
a more prominent change of the VEV.  This behavior is also seen more
clearly in {}Fig.~\ref{fig3}(b), where we show the behavior of the VEV
as a function of the electric field.  In all cases, we find that the
critical temperature, $T_c$, can be well approximately by
\begin{eqnarray}
    T_c^2& \simeq &\frac{18m^2}{\lambda},
\end{eqnarray}
which, for the model parameters chosen, it gives $T_c/M \simeq
0.424180$, in agreement with the critical temperature obtained in
{}Fig.~\ref{fig2}, when studying directly the effective potential.   As
shown in the App.~\ref{appD}, this result turns out to be a very good
approximation, with $T_c$ presenting a very weak dependence on  the
electric field, even for very strong fields, $qE \gg m^2$. 

\begin{center}
\begin{figure}[!htb]
\includegraphics[width=7.5cm]{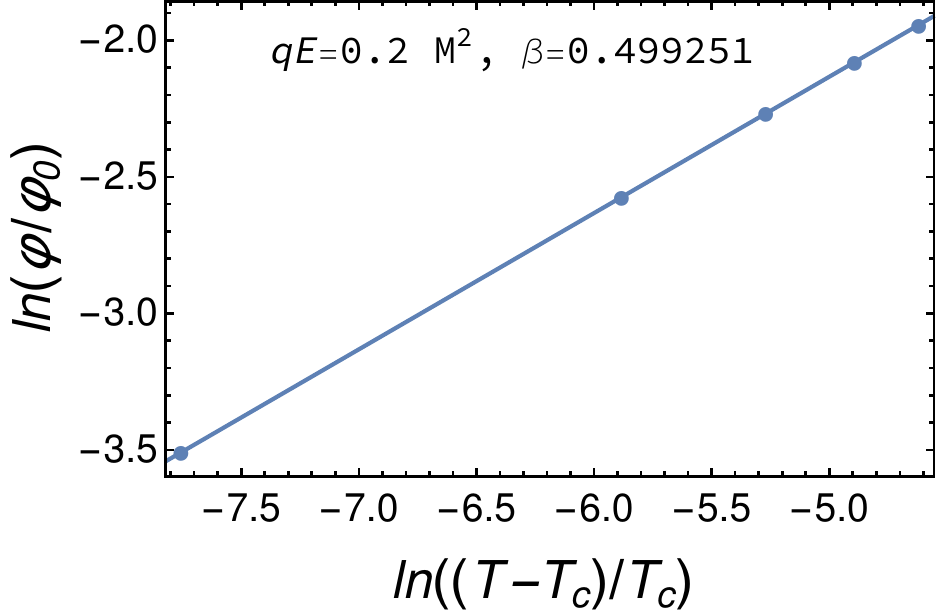}\\
\caption{Linear fitting of $\ln\left(\frac{\varphi}{\varphi_0}\right)$ as a function of $\ln\left(\frac{T-T_c}{T_c}\right)$ for different values of electric fields $qE$.}
\label{fig4}
\end{figure}
\end{center}

Next, in Fig. \ref{fig4} we explore the critical exponent, $\beta$, by performing a linear fitting of $\ln\left(\frac{\varphi}{\varphi_0}\right)$ as a function of $\ln\left(\frac{T-T_c}{T_c}\right)^{\beta}$. Due to the scale of {}Fig.~\ref{fig3}(a), it may be difficult to determine whether the second-order phase transition occurs with the critical exponent $\beta=1/2$, as expected from mean-field theory. However, the linear fit indeed shows $\beta\approx 0.5$, which is in good agreement with universality class arguments. We have numerically tested different values of the electric fields, and all results agree well with the previous statement.

%%%%%%%%%%%%%%%%%%%%%% FIG 4 %%%%%%%%%%%%%%%%%%%%%%%%%%%%%

%%%%%%%%%%%%%%%%%%%%%% FIG 5 %%%%%%%%%%%%%%%%%%%%%%%%%%%%%
\begin{center}
\begin{figure}[!htb]
\includegraphics[width=7.5cm]{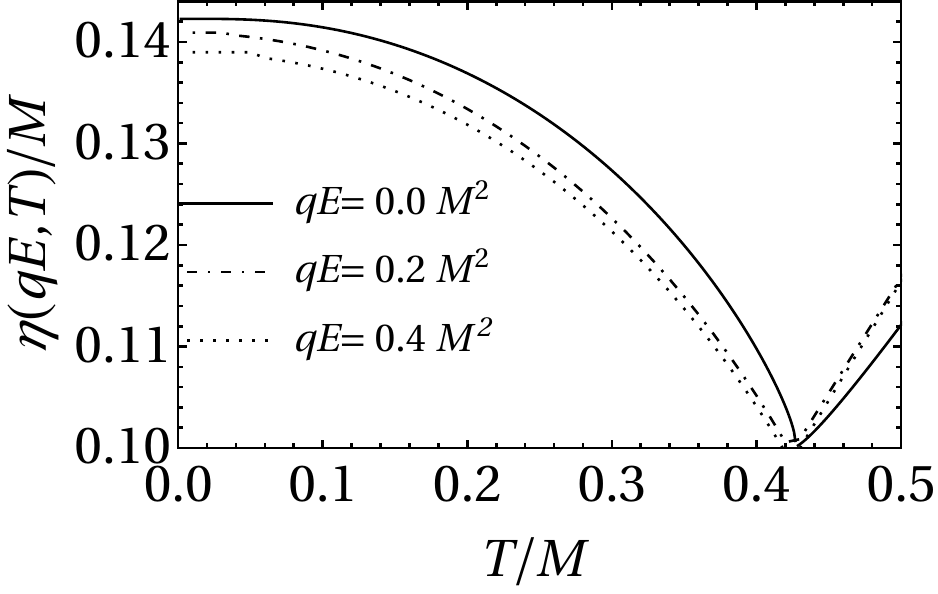}\\
\caption{The optimum OPT mass parameter $\bar\eta(qE,T)$ as a function
  of the temperature and for different values for the electric field. The model parameters chosen are $\lambda=1$ and $m=0.1M$.}
\label{fig5}
\end{figure}
\end{center}
%%%%%%%%%%%%%%%%%%%%%%%%%%%%%%%%%%%%%%%%%%%%%%%%%%%%%%%%%%

It is also useful to look at how the optimum OPT mass parameter $\bar
\eta$ changes with the temperature and with the magnitude of the
electric field.  This is shown in the {}Fig.~\ref{fig5}. {}For $qE=0$,
starting from $T=0$ up to $T=T_c$, we see the usual decrease of
$\bar\eta$ (see, e.g., Refs.~\cite{Duarte:2011ph,Silva:2023jrk}) until
it reaches the value $\bar\eta=m$ at $T=T_c$. {}For higher values of
$qE$, the values of $\bar\eta$ have a similar behavior.  {}For
$T>T_c$, $\bar\eta$ starts to grow with the temperature and the
electric field, showing an inverse behavior of the one seen in the
region with $T<T_c$.

%%%%%%%%%%%%%%%%%%%%%%%%%%%%%%%%%%%%%%%%%%%%%%%%%%%
\begin{center}
\begin{figure}[!htb]
\subfigure[]{\includegraphics[width=7.5cm]{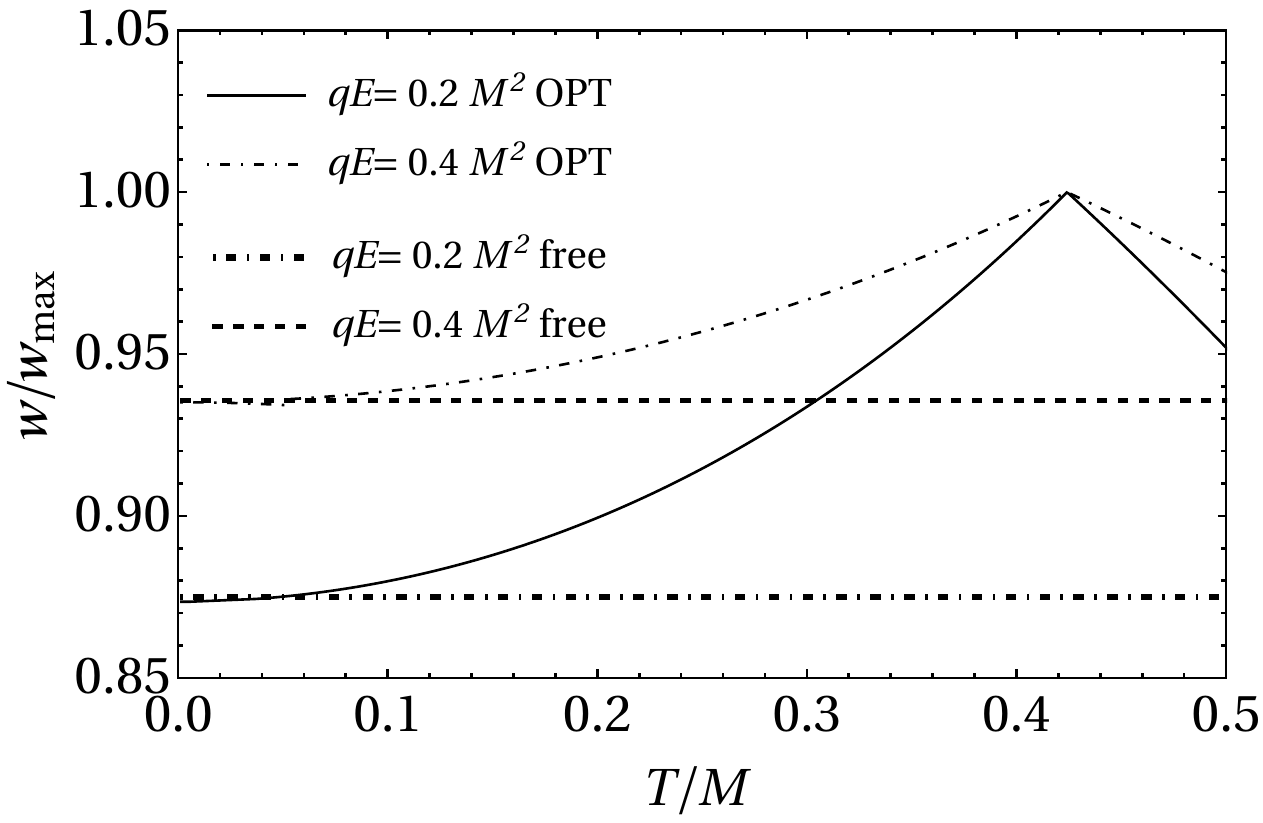}}
\subfigure[]{\includegraphics[width=7.5cm]{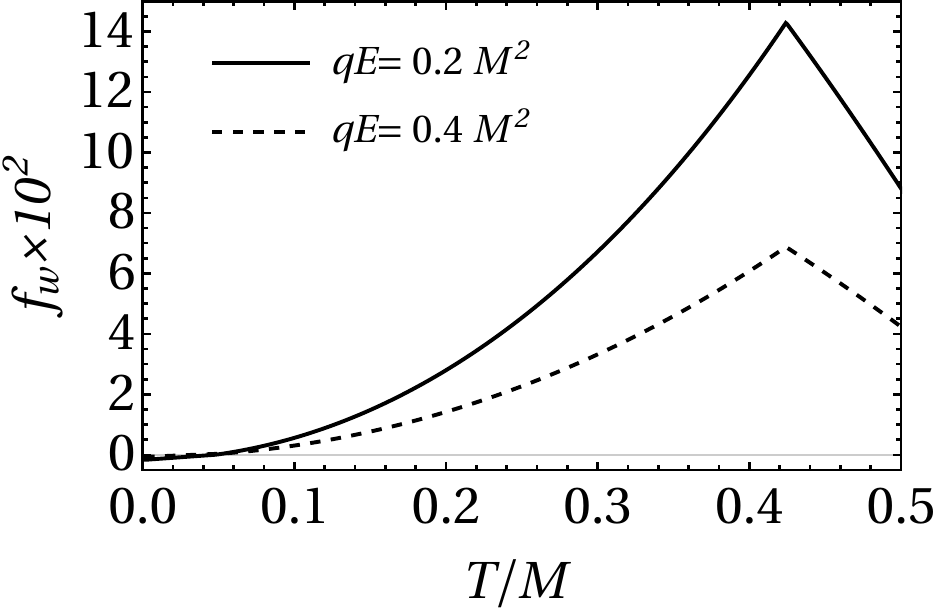}}
\caption{The vacuum persistence probability rate as a function of the
  temperature for the interaction theory (curves labeled by OPT) and
  for the free theory (panel a) and for the fractional difference,
  defined in Eq.~(\ref{fracdiff}) (panel b).}
\label{fig6}
\end{figure}
\end{center}
%%%%%%%%%%%%%%%%%%%%%%%%%%%%%%%%%%%%%%%%%%%%%%%%%%%

{}Finally, let us now investigate the temperature and electric field
dependence for the vacuum persistence probability rate, given by
Eq.~(\ref{wE}). In {}Fig.~\ref{fig6}(a) we show the results for the
interaction theory (curves labeled by OPT) and which are given  by
Eq.~(\ref{wE}). These results are also compared to the free theory
one, given by Eq.~(\ref{wfree}) (curves labeled by free). Two
representative values for the electric field are used.  In
{}Fig.~\ref{fig6}(b) we also show the fractional difference for the
rates, $f_w$, defined by
\begin{eqnarray}
f_w=\frac{w_{\text{OPT}}-w_{\text{free}}}{w_{\text{free}}}.
\label{fracdiff}
\end{eqnarray}
{}From the results shown in {}Fig.~(\ref{fig6}) we can see that the
rate $w$ peaks exactly at the critical temperature (which for the
model parameters considered, is given by 
$T_c/M \simeq 0.424180$). 
{}For
$T \ll T_c$ and for $T \gg T_c$, the rate  tends to the free scalar
QED result Eq.~(\ref{wfree}). In particular, considering the
expression for the rate Eq.~(\ref{wE}) at $T=T_c$ and using the
results from App.~\ref{appD}, we obtain that at the critical
temperature,
\begin{equation}
w(qE,T_c) \approx \frac{(q E)^2}{96 \pi},
\end{equation}
which, surprisingly, does not depend on the coupling constant, even
though away from the critical temperature, there is an explicit
dependence on the interaction in Eq.~(\ref{wE}).  We can also obtain
approximate results in the regions $T \ll T_c$ and $T \gg T_c$.  By
considering $T\ll T_c$, from Eqs.~(\ref{PMSelectric}) and
(\ref{GAPelectric}) we find that $\bar\eta^2 - 2 m^2  \approx 0$,
where we have used that $\bar \varphi \simeq \varphi_0$ for $T\ll T_c$
(see, e.g., {}Fig.~\ref{fig3}).  Hence, $\Omega^2 \approx m^2$ and
$2\pi y \approx \pi m^2/(qE)$. Also, the PMS equation
(\ref{PMSelectric}) implies that the last term on the right-hand side
in Eq.~(\ref{wE}) is negligible. As a consequence, we can conclude that  $
w_{\text{OPT}} \approx w_{\text{free}}$. This is consistent with the
result seen in {}Fig.~\ref{fig6} in the low-temperature regime. On the
other hand, for $T\gg T_c$, we are in the symmetry restored phase,
$\bar\varphi =0$. The optimum OPT parameter $\bar\eta$ in this region
grows proportional to the temperature, $\bar \eta \propto T$, as seen
from {}Fig.~\ref{fig5}, for $T>T_c$, and where the effect of the
electric field is to further enhance $\bar \eta$. In this high
temperature symmetry restored region, with also $T \gg m$, then we have the $y$ variable
in Eq.~(\ref{wE}) is much larger than $m^2/(2 qE)$ and the rate is
exponentially suppressed when compared to the free theory result
Eq.~(\ref{wfree}). Hence, $ w_{\text{OPT}} \to 0$ in this high
temperature regime.  In summary, we can conclude that the rate $w$ is
such that $w_{\rm free} \lesssim w \lesssim (q E)^2/(96 \pi)$ for $T
\leq T_c$ (symmetry broken region) 
and $0 \lesssim w \lesssim (q E)^2/(96 \pi)$ for $T\geq
T_c$ (in the symmetry restored region). 
The maximum value attained by the rate is always  $w_{\rm max}
\approx (q E)^2/(96 \pi)$, which happens at $T=T_c$. Again, this 
overall behavior is confirmed by our numerical results shown in {}Fig.~\ref{fig6}.

%%%%%%%%%%%%%%%%%%%%%%%%%%%%%%%%%%%%%%%%%%%%%%%%%%%%%%%%%%%%%
\section{Conclusions}\label{sec5}

We investigated the phase transition for symmetry restoration in a charged 
scalar field theory, considering the combined effects of temperature and 
an external constant electric field. Our study utilized the nonperturbative 
method of the optimized perturbation theory (OPT) in its first nontrivial order. 
We derived the effective potential for the model that explicitly incorporates 
the temperature and electric field dependence. To achieve this, we employed 
the Schwinger's proper time scalar propagator in the presence of a 
background electromagnetic field. Our results provide valuable insights 
into the interplay between temperature and electric fields in influencing 
the phase behavior of this system.

We numerically investigated the symmetry restoration of the model starting from 
the broken phase at low temperatures. We found that for weak electric fields, 
$qE \lesssim m^2$, the transition is first-order. As the electric field increases, 
the transition becomes second-order. We observed that increasing the 
quartic self-coupling constant lowers the threshold value of the electric field 
at which the transition changes from first to second order. Additionally, 
we studied the behavior of the critical temperature for phase transition as 
a function of the electric field. Both numerically and analytically, 
we discovered a surprisingly weak dependence on the electric field, 
which holds for both types of phase transitions. This behavior is significantly 
different from the strong dependence of the critical temperature on the magnetic 
field $B$ observed in previous studies (like in Ref.~\cite{Duarte:2011ph}) 
using the OPT at a similar order.

In addition to studying how the phase structure of the model changes
with the applied external electric field, we have also obtained the vacuum
persistence probability rate, which is defined in terms of the
imaginary part of the effective potential. We have shown, both
numerically and analytically, that at the transition point, the rate
reaches a maximum value. The maximum value found for the rate is
$w_{\rm max} \simeq (q E)^2/(96 \pi)$ which is independent of the
coupling constant of the model. We have also seen that for the
symmetry broken region, $0 \leq T \leq T_c$, the rate has a lower
bound value set by the rate of the free QED theory, given by
Eq.~(\ref{wfree}) and as the temperature increases, it increases until
reaching the value $w_{\rm max}$ at $T=T_c$. In the symmetry restored
phase, the rate decreases exponentially, vanishing asymptotically for
$T \gg T_c$.

The results presented in this paper could have significant implications 
for various fields, such as QCD topology in the presence of extreme electromagnetic fields. 
In these scenarios, the coupling between axions and photons can be calculated 
using effective models and compared with recent lattice simulations~\cite{Brandt:2023awt}. 
Understanding the behavior of meson masses as functions of magnetic fields has been 
a subject of interest, with various effective QCD models and lattice simulations 
exploring this topic~\cite{Bali:2017ian,Ding:2022tqn}. Ongoing calculations are 
investigating the effects of strong electric fields on meson masses 
and will be reported in future publications.

%%%%%%%%%%%%%%%%%%%%%%%%%%%%%%%%%%%%%%%%%%%%%%%%% 
\appendix

%%%%%%%%%%%%%%%%%%%%%%%%%%%%%%%%%%%%%%%%%%%%%%%%% 
\section{Evaluation of pure electric part of $X(\eta,T=0,qE)$}
\label{appA}

In order to obtain the pure electric field contribution for the
function $X(\eta,T,qE)$, given by Eq.~(\ref{XE}), and at $T=0$, we can
use the analytic continuation $qB\rightarrow
-iqE$~\cite{Tavares:2018poq,Tavares:2019mvq,Tavares:2023ybt} in the
pure magnetic contribution of the function $X(qB,\Omega)$ given by the
Eq.~(3.23) derived in Ref.~\cite{Duarte:2011ph},
\begin{eqnarray}
  X(qB,\Omega)&=&\frac{qB}{8\pi^2}
  \ln\left[\Gamma\left(\frac{\Omega^2}{2qB}+\frac{1}{2}\right)\right]-
  \frac{qB}{16\pi^2}\ln(2\pi)
  \nonumber\\
  &-&\frac{\Omega^2}{16\pi^2}\ln\left(\frac{M^2}{2qB}\right).
\label{A1}
\end{eqnarray}
In the first term in Eq.~(\ref{A1}), we can use the
identity~\cite{Tavares:2018poq}
\begin{eqnarray}
  \ln\Gamma(x)=-\gamma_Ex-\ln(x)+\sum_{k=1}^{\infty}\left[\frac{x}{k}-
    \ln\left(1+\frac{x}{k}\right)\right],
\end{eqnarray}
where $\gamma_E=0.57721$ is the Euler-Mascheroni constant. We can
change $x\rightarrow iy + \frac{1}{2}$, where
$y=\frac{\Omega^2}{2qE}$.  Hence, we find that
\begin{eqnarray}
  \ln\Gamma\left(iy+\frac{1}{2}\right)&=&
  -\gamma_E\left(iy+\frac{1}{2}\right)
  -\ln\left(iy+\frac{1}{2}\right)
  \nonumber\\
  &+&\sum_{k=1}^{\infty}\left[\frac{iy+\frac{1}{2}}{k}-
    \ln\left(1+\frac{iy+\frac{1}{2}}{k}\right)\right]
  \nonumber\\
  &=&-\gamma_Eiy-\frac{\gamma_E}{2}-\ln\left(iy+\frac{1}{2}\right)
  \nonumber\\
  &+&\sum_{k=1}^{\infty}\left[\frac{iy}{k}+\frac{1}{2k}-
    \ln\left(1+\frac{iy}{k}+\frac{1}{2k}\right)\right].
 \nonumber\\
\end{eqnarray}
Now, using $z=a+ib$ and that $\ln(z)$ is given by 
\begin{eqnarray}
 \ln(z)&=&\ln |a^2+b^2| + i\theta,\;\;
 \theta=\arctan\left(\frac{b}{a}\right),
\end{eqnarray}
we obtain that
\begin{eqnarray}
&& \!\!\!\!\!\! \ln\left(iy+\frac{1}{2}\right)=\ln
  \left|y^2+\frac{1}{4}\right| + i\theta,\\ &&  \!\!\!\!\!\!\!\!\!\!\!
  \ln\left(\frac{iy}{k}+\frac{1}{2k}+1\right)=\ln
  \left|\frac{y^2}{k^2}+\left(\frac{1}{2k}+1\right)^2\right| +
  i\theta_k,
\end{eqnarray}
where $\theta=\arctan(2y)$ and
$\theta_k=\arctan\left[2y/(2k+1)\right]$.  {}From these results, we
can separate $\ln\Gamma\left(iy+\frac{1}{2}\right)$ in its real and
imaginary parts as
\begin{eqnarray}
&&\!\!\!\!\!\!\!\!\!
  \Re\left[\ln\Gamma\left(iy+\frac{1}{2}\right)\right]
  =-\frac{\gamma_E}{2}-\ln\left|y^2+\frac{1}{4}\right|
  \nonumber\\
  &+&\sum_{k=1}^{\infty}\left[\frac{1}{2k}-
    \ln\left|\frac{y^2}{k^2}+\left(\frac{1}{2k}+1\right)^2\right|\right],
  \nonumber
  \\ \\ &&\Im\left[\ln\Gamma\left(iy+\frac{1}{2}\right)\right]
  =-\gamma_E y-\arctan(2y)\nonumber\\&+&\sum_{k=1}^{\infty}
  \left[\frac{y}{k}-\arctan\left(\frac{2y}{2k+1}\right)\right].
\end{eqnarray}
Hence, the first term of Eq.~(\ref{A1}), in the analytical
continuation $qB\rightarrow -iqE$, is given by
\begin{eqnarray}
    \frac{-iqE}{8\pi^2}i\Im\left[\ln\Gamma\left(iy+\frac{1}{2}\right)\right]
    =\frac{qE}{8\pi^2} \left \{-\gamma_Ey-\arctan(2y)\right.
    \nonumber\\ \left.+\sum_{k=1}^{\infty}\left[\frac{y}{k}-
      \arctan\left(\frac{2y}{2k+1}\right)\right]\right
    \}, \nonumber \\
\end{eqnarray}
while the remaining terms in Eq.~(\ref{A1}) are trivial to obtain the
analytical continuation. Thus,  the pure electric field contribution
of Eq.~(\ref{A1}) is found to be given by
\begin{eqnarray}
\lefteqn{     \Re
  X(\eta,T=0,qE)=\frac{qE}{8\pi^2}
  \left\{\vphantom{\sum_0}-\gamma_Ey-\arctan(2y)+\right.}
\nonumber\\ &&\sum_{n=1}^{\infty}\left[\frac{y}{n}
  \left.-\arctan\left(\frac{2y}{2n+1}\right)\right]\right\}+
\frac{\Omega^2}{16\pi^2}\ln\left(\frac{2qE}{M^2}\right).
\nonumber \\
\end{eqnarray}

%%%%%%%%%%%%%%%%%%%%%%%%%%%%%%%%%%%%%%%%%%%%%%%%% 
\section{Evaluation of pure electric part of $Y(\eta,T=0,qE)$}
\label{appB}

Let us now apply the same procedure as done in App.~\ref{appA} to find
the pure electric field contribution for the function $Y(\eta,T,qE)$,
given by Eq.~(\ref{YE}), and at $T=0$.  But first one notices that the
renormalized functions $X$ and $Y$ are related to each other,
\begin{eqnarray}
    \frac{\partial Y}{\partial \Omega^2}=X,
\end{eqnarray}
which can be easily verified from their general expressions, given by
Eqs.~(\ref{DX}) and (\ref{DY}).  Therefore, it is straightforward to
obtain $Y(\eta,T=0,qE)$ by
\begin{eqnarray}
   \int d\Omega^2
   X(\eta,T=0,qE)=Y(\eta,T=0,qE),
\label{intX}
\end{eqnarray}
up to an irrelevant constant to the effective potential. Then, by
simple integration, one obtains
\begin{eqnarray}
\lefteqn{\!\!\!\!\!\!\!\!\!\!\!\! \Re Y(\eta,T=0,qE)=
  \frac{\Omega^4}{32\pi^2}\ln\left(\frac{2qE}{M^2}\right)}
\nonumber
\\ &&-\frac{(qE)^2}{4\pi^2}\left\{\frac{\gamma_Ey^2}{2}+y\arctan(2y)
-\frac{1}{4}\ln\left(1+4y^2\right)\right.
\nonumber\\ &&\left.-\sum_{k=1}^{\infty}\left[\frac{y^2}{2k}-
  y\arctan\left(\frac{2y}{2k+1}\right)
  \right.\right.  \nonumber \\ &&\left.\left. +
  \frac{2k+1}{4}\ln\left(1+\frac{4y^2}{(2k+1)^2}\right)\right]\right\},
\end{eqnarray} 
which is the pure electric field part of Eq.~(\ref{Yfull}).

%%%%%%%%%%%%%%%%%%%%%%%%%%%%%%%%%%%%%%%%%%%%%%%%% 
\section{Imaginary part of functions $X(\eta,T,qE)$ and $Y(\eta,T,qE)$}
\label{appC}

By using the same techniques applied in Ref.\cite{Tavares:2018poq}, 
we can show that imaginary contribution of the thermo-electric part 
of the functions $X(\eta,qE,T)$
and $Y(\eta,T,qE)$ are zero.  Writing $X(\eta,T,qE)$ in terms of a
$T=0$  and $T\neq 0$ terms,
\begin{eqnarray}
    X(\eta,T,qE)=X_{qE}(\eta,qE)+X_{qE,T}(\eta,T,qE),
\end{eqnarray}
where $X_{qE}(\eta,qE)$ is the pure electric part and
$X_{qE,T}(\eta,T,qE)$ is thermo-electric contribution, given by
\begin{eqnarray}
    X_{qE,T}(\eta,T,qE)&=&\frac{qE}{8\pi^2}\int_0^\infty
    \frac{dt}{t\sin(t)}e^{-t\frac{\Omega^2}{qE}} \nonumber
    \\ &\times&
    \sum_{n=1}^{\infty}e^{-\frac{n^2qE}{4T^2
        |\tan(t)|}}\nonumber\\ &=&\frac{qE}{8\pi^2}
    \int_0^\infty\frac{dt\cot(t)}{t\cos(t)}e^{-t\frac{\Omega^2}{qE}}
    \nonumber \\ &\times& \sum_{n=1}^{\infty}e^{-\frac{n^2qE}{4T^2
        |\tan(t)|}},
\end{eqnarray}
where we have appropriately rewritten the terms in order to use the
following identity
\begin{eqnarray}
  \cot(t)=\frac{1}{t}+\sum_{k=1}^{\infty}\left(\frac{1}{t-k\pi}+
  \frac{1}{t+k\pi}\right).
\end{eqnarray}
The only term that contributes to the imaginary part of
$X_{qE,T}(\eta,T,qE)$ is given by
\begin{eqnarray}
    \Im X_{qE,T}(\eta,T,qE)
    &=&\frac{qE}{8\pi^2}\int_0^\infty\frac{dt}{t\cos(t)}
    e^{-t\frac{\Omega^2}{qE}}
    \nonumber\\ &\times& \sum_{n=1}^{\infty}e^{-\frac{n^2qE}{4T^2
        |\tan(t)|}}
%\nonumber\\
%&\times& 
    \sum_{k=1}^{\infty}\delta(t-k\pi) \rightarrow 0,
    \nonumber\\
\end{eqnarray}
where we have used $\lim_{\epsilon\rightarrow 0} \frac{1}{x\pm
  i\epsilon}=\text{P.V.}\frac{1}{x}\mp i\pi\delta(x)$. Then, in order
to obtain the imaginary part of both functions, we can use the result
obtained in the magnetic field system, obtained in
Ref.~\cite{Duarte:2011ph}, and using again the duality relation
$qB\rightarrow -iqE$, which gives
\begin{eqnarray}
    X(\eta,qE) &=&
    \frac{-iqE}{8\pi^2}\ln\Gamma\left(iy+\frac{1}{2}\right)+
    \frac{iqE}{16\pi^2}\ln
    (2\pi) \nonumber\\&-&
    \frac{\Omega^2}{16\pi^2}\ln\left(\frac{M^2}{-2iqE}\right).\label{IMXAPP}
\end{eqnarray}
Then, it is now easy to evaluate $\Im  X(\eta,qE)$, which is given by
\begin{eqnarray}
\Im  X(\eta,qE)\!\!  &=& \!\!-\frac{qE}{8\pi^2}\left
\{\Re\left[\ln\Gamma\left(iy+\frac{1}{2}\right)\right] -\frac{\ln
  (2\pi)}{2} + \frac{y\pi}{2}\right \}.\nonumber\\
\end{eqnarray}
By also using the identity
$\Re[\ln(iy+\frac{1}{2})]=\left[\ln\Gamma(iy+\frac{1}{2})+\ln
  \Gamma(-iy+\frac{1}{2})\right]/2$,
we can work on the first term of the  right-hand side of the previous
expression.  Then, by also using that
\begin{eqnarray}
    \frac{\ln\Gamma|(iy+\frac{1}{2})|^2}{2}
    &=&\frac{1}{2}\ln\left(\frac{\pi}{\cosh(\pi y)}\right),
\end{eqnarray}
which is obtained by analytic continuation to the complex plane of the
expression $\Gamma[x]\Gamma[1-x]=\pi(\sin(\pi x))^{-1}$, after some
algebra, we obtain
\begin{eqnarray}
    \frac{\ln\Gamma|(iy+\frac{1}{2})|^2}{2}
    &=&\frac{\ln\pi}{2}-\frac{1}{2}\ln( e^{\pi
      y})-\frac{1}{2}\ln\left(\frac{1+e^{-2\pi
        y}}{2}\right)\nonumber\\ &=&\frac{\ln
      (2\pi)}{2}-\frac{y\pi}{2}-\frac{1}{2}
    \sum_{k=1}^{\infty}\frac{(-1)^ke^{-2\pi
        y k}}{k}\nonumber\\ &=&\frac{\ln
      (2\pi)}{2}-\frac{y\pi}{2}-\frac{1}{2} \ln\left(1+e^{-2\pi
      y}\right).\nonumber\\
\end{eqnarray}
Using these results in Eq.~(\ref{IMXAPP}), we obtain
\begin{eqnarray}
    \Im X(\eta,qE) = \frac{qE}{16\pi^2}\ln\left(1+e^{-2\pi y}\right).
\label{ImX}
\end{eqnarray}
By integration in $\Omega^2$, like in Eq.~(\ref{intX}), we then obtain
\begin{eqnarray}
    \Im[Y(\eta,qE)]=\frac{(qE)^2}{16\pi^3}\text{Li}_2(-e^{-2\pi y}),
\label{ImY}
\end{eqnarray}
where we have again neglected an irrelevant constant factor when
obtaining (\ref{ImY}) and that can always be subtracted from the
effective potential and $\text{Li}_2(x)$, is the Polylogarithm
function, ${\rm Li}_2(x) = \sum_{k=1}^\infty x^k/k^2$.  The above
results show that the imaginary parts of both the functions $X$ and
$Y$ are not explicitly dependent on the temperature.

It is important to remark here that, in the context of electric fields, 
there is no clear consensus on the form or even the existence of one-loop 
thermal contributions for the imaginary part of the effective potential.
These possible thermal effects will contribute to the vacuum persistence probability 
rate~\cite{Loewe:1991mn,Elmfors:1994fw,Hallin:1994ad,Ganguly:1995mi,Ganguly:1998ys,Gies:1998vt,Gies:1999vb}. 
In particular, the worldline method has been used to derive these thermal contributions 
to the vacuum persistence probability rate~\cite{Medina:2015qzc,Gould:2017fve,Korwar:2018euc}.
These studies have shown that the thermal contributions become relevant (compared to the
$T=0$ result) at a threshold temperature $T_{CW}=\frac{eE}{2 \Omega}$. Below $T_{CW}$
the thermal bath's energy is insufficient to create 
a charged particle-antiparticle pair accelerated by the electric field over their Compton wavelength. 
In this regime, thermal effects are negligible, and charged particle-antiparticle 
pairs are produced only through quantum processes. 
All the numerical results presented in Sec.~\ref{sec4} focus on the region where 
these possible thermal contributions to the effective potential and vacuum persistence probability rate 
can be ignored, i.e., we focus on the regime $T<T_{CW}$.

%%%%%%%%%%%%%%%%%%%%%%%%%%%%%%%%%%%%%%%%%%%%%%%%% 
\section{Independence of the critical temperature with the electric fields}
\label{appD}

Let us now demonstrate that the critical temperature is indeed very
weakly dependent on the electric field.  To evaluate the critical
temperature, we consider the situation where $\bar\varphi=0$ and,
therefore, $\Omega=0$,  since $\bar \eta^2=m^2$ and as demonstrated in
Ref.~\cite{Duarte:2011ph}. This is an approximation of
Eq.~(\ref{PMSelectric}), since we are considering the second
contribution very small. The second contribution term in
Eq.~(\ref{PMSelectric}), and that involves $\Im X$, can be neglected,
which  is justified by the numerical results we have obtained in
Sec.~\ref{sec4}, and implies that $T_c(qE)\approx T_c(0)$. Then,
\begin{eqnarray}
    \frac{6m^2}{\lambda}-4X(\bar\eta,T_c,qE)\approx 0,\label{XD1}
    \end{eqnarray}
where
\begin{eqnarray}
  X(\bar\eta,T_c,qE)&=&\frac{qE}{8\pi^2}\int_0^{\infty}
  \frac{dt}{t\sin(t)}\sum_{n=1}^{\infty}e^{-\frac{n^2qE}{4T_c^2|\tan(t)|}}.
\nonumber \\
\label{XD2}
\end{eqnarray}
{}From a change of variables, we can also write Eq.~(\ref{XD2}) as
\begin{eqnarray}
   X(\bar\eta,T_c,qE)&=&\sum_{n=1}^{\infty}\frac{qE}{8\pi^2}
\int_0^{\infty}\frac{d\tau}{\sqrt{1+\tau^2}}e^{-\frac{n^2qE}{4T_c^2}\tau}
   \nonumber\\ &&\times\frac{4}{\pi}\sum_{k=0}^{\infty}
\left[\frac{(-1)^k(2k+1)}{(2k+1)^2-\left(\frac{2}{\pi}\tan^{-1}\tau\right)^2}\right].
   \nonumber \\
\label{XD4}
\end{eqnarray}
Now, we can make use of the identity~\cite{prudnikov1986integrals}, 
\begin{eqnarray}
    \sum_{k=0}^{\infty}\left[\frac{(-1)^k(2k+1)}{(2k+1)^2-a^2}\right]=\frac{\pi}{4}\sec\left(\frac{\pi
      a}{2}\right),
\end{eqnarray}
as well as the property $\sec(\tan^{-1}\tau)=\sqrt{1+\tau^2}$. Then,
applying these results in Eq.~(\ref{XD4}), we obtain
\begin{eqnarray}
X(\bar\eta,T_c,qE)&=&\frac{qE}{8\pi^2}\sum_{n=1}^{\infty}\int_0^{\infty}d\tau
e^{-\frac{n^2qE}{4T_c^2}\tau},
\label{XD5}
\end{eqnarray}
which can be easily integrated, resulting in
\begin{eqnarray}
X(\bar\eta,T_c,qE)&=&\frac{T_c^2}{12},
\end{eqnarray}
which is exactly the same result as obtained in
Ref.~\cite{Duarte:2011ph} in the case of $qE=0$.  When we apply the
last result in Eq.~(\ref{XD1}), we obtain
\begin{eqnarray}
    T_c^2=\frac{18m^2}{\lambda},
\end{eqnarray}
and which is in full agreement with the numerical results obtained in
Sec.~\ref{sec4}.

%%%%%%%%%%%%%%%%%%%%%%%%%%%%%%%%%%%%%%%%%%%%%%%%% 
\section*{Acknowledgments}

This work was partially supported by Conselho Nacional de
Desenvolvimento Cient\'ifico  e Tecno\-l\'o\-gico  (CNPq),
312032/2023-4 (R.L.S.F.),  304518/2019-0 (S.S.A.), 307286/2021-5
(R.O.R.); Fundação Carlos Chagas Filho de Amparo à Pesquisa do  Estado
do Rio de Janeiro (FAPERJ), Grant No.SEI-260003/019544/2022 (W.R.T),
Grant No. E-26/201.150/2021 (R.O.R.);  Funda\c{c}\~ao de Amparo \`a
Pesquisa do Estado do Rio Grande do Sul (FAPERGS), Grants
Nos. 19/2551- 0000690-0  and 19/2551-0001948-3 (R.L.S.F.); The work is
also part of the  project Instituto Nacional de Ci\^encia e Tecnologia
- F\'isica Nuclear e Aplica\c{c}\~oes  (INCT - FNA), Grant
No. 464898/2014-5. 

%%%%%%%%%%%%%%%%%%%%%%%%%%%%%%%%%%%%%%%%%%%%%%%%%

\end{document}